\documentclass[fleqn,iicol]{sn-jnl}

\usepackage{graphicx}%
\usepackage{multirow}%
\usepackage{amsmath,amssymb,amsfonts}%
\usepackage{amsthm}%
\usepackage{mathrsfs}%
\usepackage[title]{appendix}%
\usepackage{xcolor}%
\usepackage{textcomp}%
\usepackage{manyfoot}%
\usepackage{booktabs}%
\usepackage{algorithm}%
\usepackage{algorithmicx}%
\usepackage{algpseudocode}%
\usepackage{listings}%
\usepackage{dsfont}
\usepackage{color}
\usepackage{longtable}
\usepackage{verbatim} 
\usepackage{hyperref}
\usepackage{subcaption}
\usepackage{graphicx}
\usepackage{nicefrac}
\DeclareCaptionLabelFormat{nocaption}{}

\begin{document}

\title[Looking for optimal materials for WGM applications at $2\,\mu$m]{Looking for optimal materials for whispering gallery modes applications at the $2\,\mu$m window}                               
\author*[1]{\fnm{Lorena} \sur{Velazquez-Ibarra}} 
\email{lorenav@fisica.ugto.mx}

\author[1]{\fnm{Juan} \sur{Barranco}}
\email{jbarranc@fisica.ugto.mx}

\affil*[1]{\orgdiv{Divisi\'on de Ciencias e Ingenier\'ias},  \orgname{Universidad de Guanajuato, Campus Le\'on},
 \postcode{37150}, \city{Le\'on}, \state{Guanajuato}, \country{M\'exico}}


\abstract
{The diverse applications of whispering gallery modes in spherical microresonators are strongly related to the sphere size and material composition. Their design should therefore be optimized to ensure that parameters such as the quality factor and the free spectral range are maximized. Because of the imminent capacity crisis of the optical communication systems operating at the 1550\,nm wavelength regime, it is time to explore optical communications at the 2\,$\mu\mbox{m}$ wavelength window. In this work, we analytically investigate key resonator parameters -~quality factor and free spectral range~- as a function of wavelength, aiming to establish a methodology to help identify optimal materials for whispering gallery mode sensors, with special attention at the 2\,$\mu\mbox{m}$ wavelength window. Specifically, we examine three materials: fused silica, AsSe chalcogenide glass and calcium fluoride, and we perform a comparison between them in order to identify the region in the parameter space of resonant wavelengths and sphere radius, $(\lambda_R,R$), where the WGM resonators are optimal at wavelengths $1.8\,\mu\mbox{m}<\lambda<2.1\,\mu\mbox{m}$.}




\maketitle

\section{Introduction}

Optical communications systems are approaching their saturation limits at the telecommunication wavelength bands ($\sim 1250$\,nm$-1650$\,nm). 
Motivated by this necessity of extending photonics applications beyond the telecom wavelength range, 
explorations of the 2\,$\mu$m window are of current interest~\cite{Gunning:19,Wang:23,2micras_entanglement}. 
This new window opens a new world for extremely interesting sensing and biological applications. For instance, the monitoring associated with CO$_2$~\cite{Wang:18}, or the analysis of blood glucose~\cite{Alexeeva09} will be greatly improved at wavelengths around this $2\,\mu$m window. In an orthogonal application, even for gravitational wave observatories, this wavelength band is of current interest~\cite{Wills:18}. 
In this spirit, other resonators operating at $2\,\mu$m could offer new opportunities not yet explored. Some of the highest quality optical resonators are achieved with whispering gallery mode (WGM) microresonators. In WGM optical resonators, the light circulates along their concave inner surface and, if the material dielectric properties allow minimal reflection losses and low material absorption, these resonators can reach exceptionally high quality factors ($Q$-factors). Furthermore, they can be designed to have narrow spectral linewidths, and very high sensitivity to small changes in refractive index, making them a valuable tool for a wide range of photonics applications. Most of those applications have been centered at the optical band~\cite{Righini:16,Jiang:20,Barman:24} mainly because of the availability of laser sources.
With the imminent extension of optical communications at the $2\,\mu$m window, new lasers around these wavelengths are already available~\cite{Yin:19,Latkowski:16,laser2micras}. Thus, it is a good time to study the properties of WGM microresonators at this wavelength window. For previous work exploring applications of WGM at the 2\,$\mu$m window for biosensors and thermal sensors see ~\cite{Righini:16,Yang:17,Jiang:20}, or for specialty light sources like white-light sources with tailored emission or Brillouin lasers with $2\,\mu$m emission see
~\cite{Barman:24,Pathak:24}. 

In the present work, we explore different materials for WGM in microspheres beyond the 1550\,nm region, aiming to optimize certain properties such as: the $Q$-factor, the free spectral range (FSR), and the sensitivity of the FSR to changes in the sphere refractive index, to aid in the design of microresonators for different applications. These properties depend on the optical material and on the size of the sphere.

For definiteness, we study and compare the following materials: 

\begin{enumerate}
\item Fused silica (SiO$_2$): it is a widely used material in optical systems due to some remarkable optical properties, such as high purity and broad transparency window (extending from UV to NIR); 

\item Calcium fluoride (CaF$_2$): it has a long history as a semiconductor material, and it has also been widely used in optics because of its transparency region from UV to IR with minimal absorption. The second laser ever made used uranium doped calcium fluoride to emit at 2.5\,$\mu$m~\cite{Sorokin:60}, and similarly, stimulated emission into optical WGM in CaF$_2$:Sm$^{++}$ spheres was observed shortly after~\cite{Garrett:61}. Comparably to silica, calcium fluoride is also relatively abundant.

\item AsSe chalcogenide: it is a good prospect for applications that require IR transmission due to its wide transparency from visible to IR. Additionally, chalcogenide glasses are typically highly nonlinear~\cite{Zakery:03}, making them a useful choice for nonlinear and quantum applications~\cite{Strekalov:16}.
\end{enumerate}

A comprehensive study of optical materials for their use in WGM resonators is beyond the scope of the present work; nevertheless, the proposed methodology can be applied to different dielectric materials.

Our analysis reveals that in the 1.8\,$\mu$m~-~2.1\,$\mu$m wavelength range, SiO$_2$ has the highest $Q$-factor and the highest sensitivity of the FSR under changes of the refractive index, when compared to AsSe chalcogenide and CaF$_2$, if the radius of the sphere is $R>20\,\mu$m. For smaller radii, the $Q$-factor decreases in this window. Thus, for WGM resonators, silica exhibits optimal characteristics for sensing based on mode shifting sensing. This extends silica applications beyond optical logics and signal processing, where silicon is explored in the same window as well. 

In order to validate our conclusion, we have organized the paper as follows: in Section~\ref{sec:optprop} we present the optical properties of the studied materials that are relevant for the calculation of the WGM resonant wavelengths; in Section~\ref{sec:compute} we present the calculation of the resonant wavelengths; in section~\ref{sec:WGMprop} we present relevant formulae for the computation of Q-factor and FSR. In section~\ref{sec:disc} we discuss our results and conclusions. 

\section{Optical properties of silica, calcium fluoride and AsSe chalcogenide}\label{sec:optprop}

\begin{table*}
\caption{Sellmeier coefficients for bulk-fused silica~\cite{Agrawal}, calcium fluoride~\cite{Li:80}, and AsSe chalcogenide~\cite{Dantanarayana:14}.}
\begin{tabular*}{\textwidth}{@{\extracolsep\fill}l lllllll}
\toprule 
Material & $a_0$ & $a_1$ & $a_2$ & $a_3$ & $b_1\,[\mu$m] &  $b_2\,[\mu$m] &  $b_3\,[\mu$m] \\\midrule \midrule
Fused silica & $1.0$ & $0.6961663$ & $0.4079426$ 
 & $0.8974794$ & $0.0684043$  & $0.1162414$  &  $9.896161$ \\ \midrule
 Calcium fluoride & $1.33973$ & $0.69913$ & $0.11994$ & $4.35181$ & $0.09374$  & $21.18$ & $38.46$ \\ \midrule
 AsSe chalcogenide & $3.5611$ & $1.9098$ & $0.89565$  & $2.1738$ & $0.47350$ 
 & $41.383$ & $0.31202$ \\ \midrule
\end{tabular*}
\label{tab:sell}
\end{table*}
 In a simplified picture of WGM spherical resonators surrounded by air, a ray of incoming light is trapped inside the sphere hitting its surface if the angle of incidence is bigger than a critical angle, $i_c$, given in terms of the refractive index of the sphere material, $n(\lambda)$, as $i_c=\sin^{-1}(1/n(\lambda))$, so that total internal reflection occurs.
 Thus, the important material optical property involved in the theoretical computation of the resonant wavelengths $\lambda_R$ of a WGM is the refractive index, which in dispersive materials is a function of wavelength, $n(\lambda)$. 

\begin{figure}[ht]
\captionsetup[subfigure]{labelformat=nocaption}
\centering
\includegraphics[width=7.6cm]{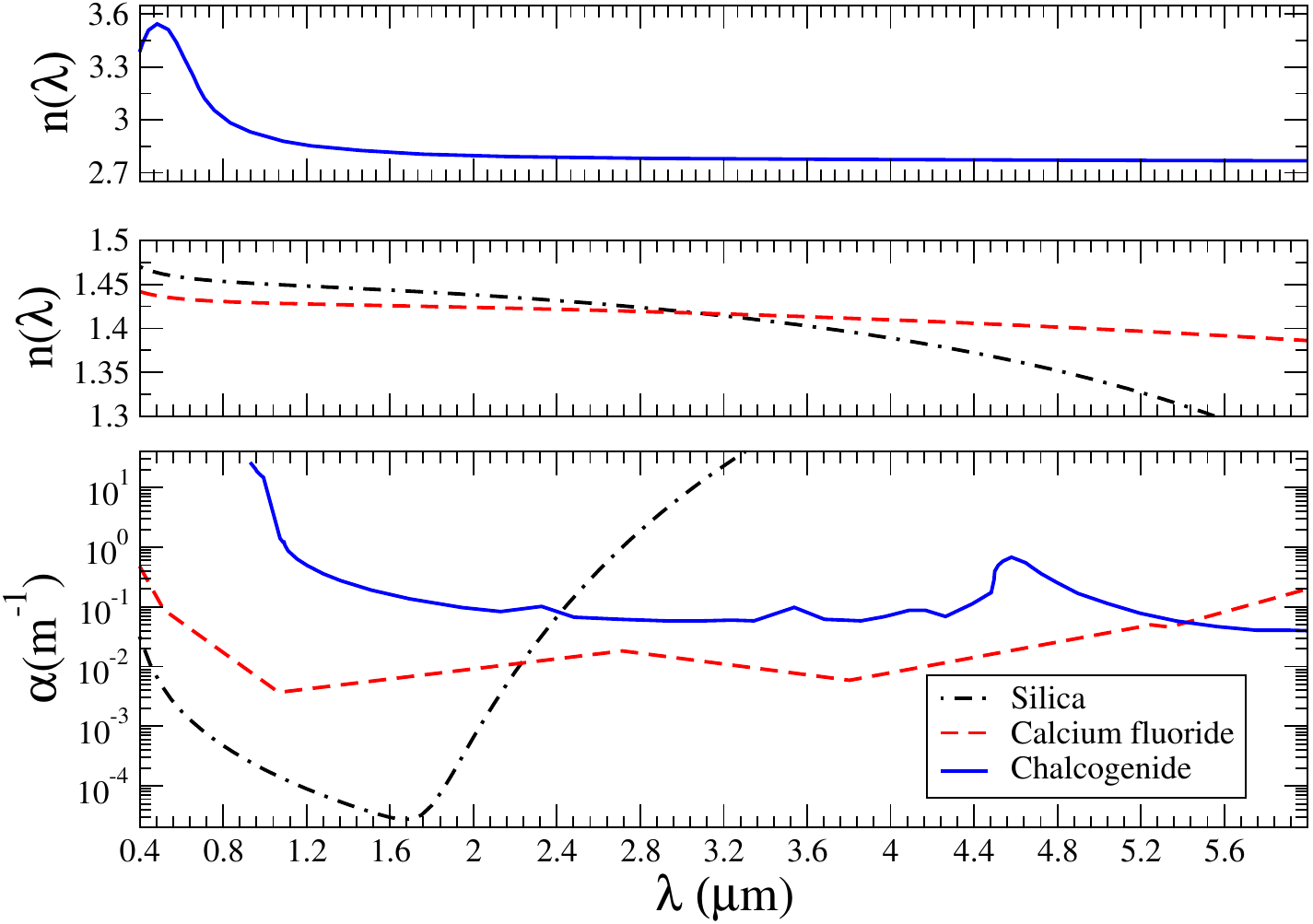}%
    \begin{subfigure}{0\linewidth}
    \caption{}\label{fig:nalphaa}
    \end{subfigure}
    \begin{subfigure}{0\linewidth}
    \caption{}\label{fig:nalphab}
    \end{subfigure}
    \begin{subfigure}{0\linewidth}
    \caption{}\label{fig:nalphac}
    \end{subfigure}
 \caption{(a) and (b) Refractive index, $n(\lambda)$, and (c)~absorption coefficient, $\alpha(\lambda)$, for: silica (black dot-dashed line), calcium fluoride (red dashed line) and AsSe chalcogenide (blue solid line).}
    \label{fig:nalpha}
\end{figure}

As usual, the refractive index as a function of wavelength is given by the Sellmeier equation~\cite{Sellmeier1872},
\begin{equation}
n^2(\lambda)=a_0+\sum_{j=1}^3 \frac{a_j \lambda^2}{\lambda^2-b_j^2},
\label{sellmeier}
\end{equation}
where the Sellmeier coefficients $a_j$ and $b_j$ for the materials chosen for our study are given in Table~\ref{tab:sell}. Data for bulk-fused silica has been measured with high accuracy and reported in~\cite{Malitson:65,Agrawal}. For calcium fluoride, data has been compiled from~\cite{Li:80}, and for AsSe chalcogenide from~\cite{Dantanarayana:14}.
Fig.~\ref{fig:nalphaa} is a plot of the refractive index $n(\lambda)$ of AsSe chalcogenide, while Fig.~\ref{fig:nalphab} shows the refractive index for both fused silica and calcium fluoride. It is clear that SiO$_2$ and CaF$_2$ have very similar indices of refraction for wavelengths $\lambda<4\,\mu$m. 
On the other hand, the refractive index for AsSe chalcogenide is higher than the other two materials considered in this study.

In addition to the condition of resonance, an important characteristic of a good resonator is that the confined modes survive a relatively long time inside the sphere before dissipating due to losses. This lifetime depends directly on the absorption coefficient, $\alpha(\lambda)$, of the material where the WGM is propagating. Thus, in order to analyze if the materials under study are adequate for WGM resonators, a good knowledge of their absorption coefficients is needed. 

 The absorption coefficient for fused silica is well known and has been measured with high accuracy. A functional expression has been extrapolated from experimental data as follows~\cite{Gorodetsky:00}: 

\begin{eqnarray}
    \nonumber \alpha(\lambda)&=&\biggr(\frac{0.7\,\mu\mbox{m}^4}{\lambda^4}+1.1\times10^{-3}\exp{\frac{4.6\,\mu\mbox{m}}{\lambda}}\\ 
    &&+4\times10^{12}\exp{\frac{-56\,\mu\mbox{m}}{\lambda}}\biggr)\frac{\mbox{dB}}{\mbox{km}}\,. \label{eq:alpha}
\end{eqnarray}

Nevertheless, it is not always possible to obtain a unified description of the absorption coefficient for all the materials. This is because the absorption $\alpha(\lambda)$ varies depending on several factors, like the presence of impurities in the material or the fabrication method. 

For instance, although extensive literature exists on experimental measurements of the absorption coefficient of calcium fluoride across several wavelengths, data for the visible region remains limited. For definitiveness, in this work we use data reported in~\cite{Li:80:2, corning}, which is summarized in Table~\ref{tab:alfaCaChal}. 

The available data on the absorption coefficient for AsSe chalcogenide is more limited. For the purposes of this work, we have taken data reported in~\cite{Ballato:17, Boudebs:04}, and summarized in Table~\ref{tab:alfaCaChal}. 

Fig.~\ref{fig:nalphac} shows the absorption coefficient $\alpha(\lambda)$ for the three materials under study.  Note that the absorption coefficient for silica reaches a minimum value around the telecommunications wavelength $\lambda=1550$\,nm. We can observe that $\alpha(\lambda)$ for silica is higher than that for calcium fluoride and for AsSe chalcogenide for $\lambda>2.4\,\mu$m. This roughly explains why CaF$_2$ and AsSe are expected to show a better performance than silica for wavelengths in the NIR region. 

\begin{table}
\caption{Absorption coefficient $\alpha$ for calcium fluoride~\cite{Li:80:2,corning} and AsSe chalcogenide~\cite{Ballato:17,Boudebs:04}.}
\begin{tabular}{lll}
\toprule 
& Calcium fluoride  & AsSe chalcogenide   \\ \midrule 
$\lambda\,(\mu$m)    & \multicolumn{2}{@{}c@{}}{$\alpha$\,($m^{-1}$) }\\\midrule \midrule
$0.167$ & $16.82172$ & $-$ \\ \midrule
$0.185$ & $10.47912$ & $-$\\ \midrule
$0.208$ & $7.65948$ & $-$\\ \midrule
$0.515$ & $0.08937$ & $-$\\ \midrule
$1.066$ & $0.0037$ & $2.507753$\\ \midrule
$2.714$ & $0.01829$ & $0.0514$\\ \midrule
$3.804$ & $0.0059$ & $0.0538$\\ \midrule
$5.25$ & $0.05$  & $0.0692$ \\ \midrule
$5.338$ & $0.04587$ & $0.00607$ \\ \midrule
$6.250 $ & $0.35$ & $0.0371$\\ \midrule
$7.690 $ & $11$ & $0.052$\\ \midrule
$8.690 $ & $48$ & $0.0857$\\ \midrule
$9.090 $ & $65$ & $0.125$\\ \midrule
$10.600$ & $350$ & $4.3$\\ \midrule
\end{tabular}
\label{tab:alfaCaChal}
\end{table}

\section{WGM resonances computation}\label{sec:compute}

Whispering gallery modes in a sphere can be understood as electromagnetic modes that propagate along the surface of a circular cavity by repeated total internal reflection. One of the simplest three-dimensional geometries where WGM can be found is that of a sphere. We consider in this work a sphere of radius $R$, made of a dielectric material with refractive index $n(\lambda)$ and surrounded by air, so the external refractive index is set to unity. 

The electromagnetic problem consists in finding the solutions to the Helmholtz equation, with boundary conditions of continuity of the tangential components of the electromagnetic fields $\vec E$ and $\vec B$ at the surface of the sphere. There are two different polarizations: either transverse electric (TE) if the electric is parallel to the surface,  or transverse magnetic (TM) if magnetic field is parallel to the surface of the sphere, respectively. In the present work, we focus on TE modes. A similar analysis can be carried out for TM modes, and the results read out analogously. 

The fields for TE polarization modes are expressed as:

\begin{equation}
\mathbf{E}_{\ell}^m(\mathbf{r})=A\frac{F_{\ell}(r)}{k_0(\lambda)r}\mathbf{X}_{\ell}^m (\theta,\phi)  
\end{equation}

\begin{eqnarray}
\mathbf{B}_{\ell}^m(\mathbf{r})&=&\frac{A}{ic}\biggr(\frac{F'_{\ell}(r)}{k_0^2(\lambda)r}\mathbf{Y}_{\ell}^m (\theta,\phi) \nonumber \\ &+&\sqrt{\ell(\ell+1)}\frac{F_{\ell}(r)}{k_0^2(\lambda)r^2}\mathbf{Z}_{\ell}^m(\theta,\phi) \biggr)
\end{eqnarray}

\noindent{where $k_0(\lambda)=2\pi/\lambda=\omega/c$ is the free-space wavenumber, $c$ is the speed of light in vacuum, $\mathbf{X}_{\ell}^m (\theta,\phi)$, $\mathbf{Y}_{\ell}^m (\theta,\phi)$ and $\mathbf{Z}_{\ell}^m (\theta,\phi)$ are the vectorial spherical harmonics defined in terms of the scalar spherical harmonics $Y_{\ell}^m$ as}

\begin{eqnarray}
    \nonumber \mathbf{X}_{\ell}^m&=&
    \frac{1}{\sqrt{\ell(\ell+1)}}\nabla   Y_{\ell}^m\times\mathbf{r}\,,\\
    \mathbf{Y}_{\ell}^m&=&\frac{1}{\sqrt{\ell(\ell+1)}}r\nabla
    Y_{\ell}^m\,,\\
    \nonumber \mathbf{Z}_{\ell}^m&=&Y_{\ell}^m\hat{r}.
\end{eqnarray}

The radial function $F_{\ell}(r)$ is a linear combination of the Riccati-Bessel functions $ \psi_{\ell}(x)=x j_{\ell}(x)$ and $\chi_{\ell}(x)=x y_{\ell}(x)$, where, $j_{\ell}(x)$ and $y_{\ell}(x)$ are the spherical Bessel and Neumann functions, respectively.
By demanding regularity at $r=0$, the radial function inside the sphere only has the contribution of the spherical Bessel function since the Neumann function diverges at $r=0$, thus,

\begin{equation}
    F_{\ell}^{\mathrm{in}}(r)=
        \psi_{\ell} (k_0(\lambda) n(\lambda)r) \,,
    \label{eq:fradi}
\end{equation}
for $0\le r\le R$. On the other hand, the solution of the radial function for the exterior of the sphere is

 \begin{equation}
    F_{\ell}^{\mathrm{out}}(r)=  C \psi_{\ell} (k_0(\lambda)r) + D \chi_{\ell} (k_0(\lambda)r)\,, 
    \label{eq:frado}
\end{equation}
for $r>R$.

The condition of continuity of the tangential components of TE modes implies conditions at $r=R$ for the radial function $F_\ell(r)$ that read as follows: 

\begin{eqnarray}
     F_{\ell}^{\mathrm{in}}(r=R)&=& F_{\ell}^{\mathrm{out}}(r=R)\nonumber \\
    n(\lambda)F_{\ell}^{\mathrm{' in}}(r=R)&=& F_{\ell}^{\mathrm{' out}}(r=R).
    \label{eq:sys}
\end{eqnarray}

From this point onwards, we drop the explicit wavelength dependence from the wavenumber, $k_0$, and the refractive index, $n$, to simplify the notation. Solving Eq.~\eqref{eq:sys} to find the coefficients $C$ and $D$ in Eq.~\eqref{eq:frado} results in solving the following eigenvalue equation for $\lambda$, 

\begin{equation}
    n\frac{j'_{\ell}(k_0nR)}{j_{\ell} (k_0 nR)}=\frac{
    y'_{\ell}(k_0R)}{y_{\ell} (k_0R)}.
    \label{eq:complete}
\end{equation}

The eigenvalue equation for the resonant wavelengths given by Eq.~\eqref{eq:complete} is usually expressed in terms of the cylindrical Bessel functions. Namely, 

\begin{equation}
    n\frac{J'_{\nu}(k_0nR)}{J_{\nu} (k_0nR)}=\frac{
    Y'_{\nu}(k_0R)}{Y_{\nu} (k_0R)},
    \label{eq:completecyl}
\end{equation}
where $J_\nu$ and $Y_{{\nu}}$ are the cylindrical Bessel and cylindrical Neumann functions, respectively, of half integer order 

\begin{equation}
    \nu~=l+\frac{1}{2}\,.
    \label{eq:nu}
\end{equation}

Henceforth, since $\nu$ depends on $\ell$, we denote $\nu$ as $\nu_\ell$.

\begin{figure}[ht]
\includegraphics[width=0.49\textwidth]{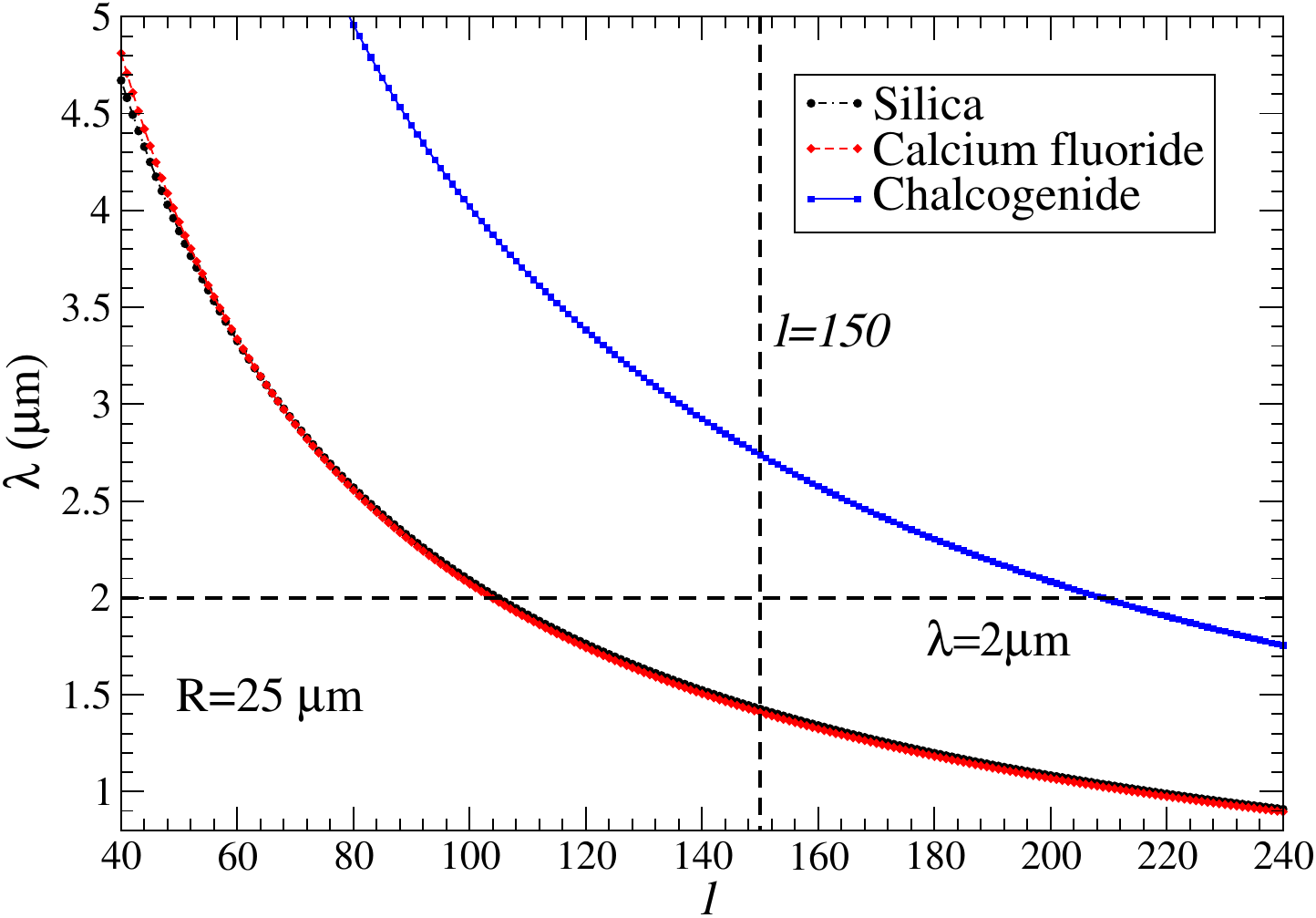}
\caption{Resonant wavelengths of the first radial mode as a function of the azimuthal number, for a fixed radius $R~=~25\,\mu$m, for the three materials under consideration.}
\label{fig:lamda_res}
\end{figure}

There is an infinite number of solutions to Eq.~\eqref{eq:complete}, even for a  fixed value of $\ell$. We label each of this solutions by the index $q$.  The value $q=1$ corresponds to the first root and, correspondingly, to the first radial mode of the electromagnetic field associated with the WGM in which the radial function $F_\ell(r)$ has zero nodes. For $q>1$, $F_\ell$ will have $q-1$ nodes. Hence, we denote the resonant wavelengths as $\lambda_{TE,\ell}^{R,q}$.
 
Solving Eq.~\eqref{eq:complete} can be time consuming. For this reason, approximated solutions have been found~\cite{Lam:92,Schiller:93}. 
To a first order in powers of $\nu_\ell^{-\ell/3}$, $\lambda_{TE,\ell}^{R,q}$ can be computed as:

\begin{eqnarray}
    \lambda_{TE,\nu_\ell}^{R,q}&=&2\pi  nR\biggr[\nu_\ell 
    -t^0_q\left(\frac{\nu_\ell}{2}\right)^{1/3} \nonumber\\
    &-&\frac{n}{(n^2-1)^{1/2}}+\frac{3}{20} (t^q_n)^2\left(\frac{\nu_\ell}{2}\right)^{-1/3} \nonumber \\&+& O(\nu_\ell^{-1})\biggr]^{-1}\,.
    \label{eq:aproximated}
\end{eqnarray}

Here, $t^0_q$ is the $q$-th zero of the Airy function. The value $q=1$ corresponds to the first zero and, correspondingly, to the first radial mode. 
Since in dispersive materials the refractive index is not constant with wavelength, even this approximate expansion needs to be solved numerically. For large values of $\ell$, the differences between the resonant wavelengths computed via Eq.~\eqref{eq:aproximated} compared to via the exact eigenvalue Eq.~\eqref{eq:complete} are smaller that $0.1\%$ \cite{Velazquez:24}. Thus, for the purposes of this work, we solve Eq.~\eqref{eq:aproximated} without important changes in the conclusions. 

\subsection{WGM in Silica, Calcium fluoride and Chalcogenide}

\begin{figure*}[ht]
\captionsetup[subfigure]{labelformat=nocaption}
\centering
\includegraphics[width=0.98\textwidth]{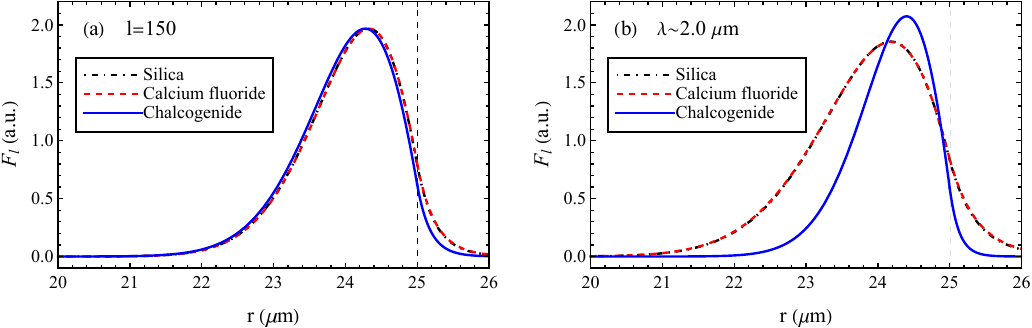}%
    \begin{subfigure}{0\linewidth}
    \caption{}\label{fig:fra}
    \end{subfigure}
    \begin{subfigure}{0\linewidth}
    \caption{}\label{fig:frb}
    \end{subfigure}
   \centering
 \caption{Radial TE field intensity of the fundamental radial mode in a microsphere with $R=25$~$\mu$m and: (a) azimuthal mode $\ell$=150 ($\lambda_{SiO_2} = 1.4264\,\mu$m, $\lambda_{CaF_2} = 1.4080\,\mu$m, $\lambda_{AsSe} = 2.7379\,\mu$m); (b) resonant wavelength around 2~$\mu$m ($\ell_{SiO_2} = 104$, $\ell_{CaF_2} = 103$, $\ell_{AsSe} = 208$).}
    \label{fig:fr}
\end{figure*}

Based on the discussion presented in the previous sections, we now compute the WGM resonant wavelengths in a sphere of radius $R$. The information needed for solving Eq.~\eqref{eq:aproximated} for each material is the refractive index $n(\lambda)$.
In order to start the comparative analysis of the different materials under consideration  (SiO$_2$, CaF$_2$ and AsSe chalcogenide), we compute the resonant wavelengths with the refractive index given by the Sellmeier Eq.~\eqref{sellmeier} 
with the coefficients $a_i$ and $b_i$ given in Table~\ref{tab:sell} corresponding to each material.

To find the real part of the resonant wavelengths of the WGM, Eq.~\eqref{eq:aproximated} is solved numerically using a Newton method for the first radial order, $q=1$, and different values of the azimuthal number, $l$. The analysis for different values of $q$ can be carried out analogously; the results and the conclusions are similar to those obtained for $q=1$. 
 
The resonant wavelengths for the first radial mode ($q=1$) in a sphere of radius $R=25\,\mu$m are shown in Fig.~\ref{fig:lamda_res}, for each of the three materials as a function of $\ell$. 

Some conclusions can be inferred from Fig.~\ref{fig:lamda_res}: \\
The resonant wavelengths $\lambda_{TE,\nu_\ell}^{R,1}$ numerical values for silica and calcium fluoride differ only by around $1\%$  for the same value of $\ell$. This is a direct consequence of the similarities between the refractive indices of SiO$_2$ and CaF$_2$.

On the other hand, the refractive index of AsSe chalcogenide is significantly higher than the refractive indices of silica and calcium fluoride, therefore, the resonant wavelengths $\lambda_{TE,\nu_\ell}^{R,1}$ for AsSe are longer, almost twice the value of the resonant wavelengths for silica and calcium for the same $\ell$ and radius. 

In order to better understand what is physically happening, we have plotted the radial part of the electromagnetic fields given by Eqs.~\eqref{eq:fradi} and \eqref{eq:frado}, for the resonant wavelengths $\lambda_{TE,\nu_\ell}^{R,1}$ previously computed.
Fig.~\ref{fig:fra} shows the plot of the radial functions for a sphere of radius $R=25\,\mu$m  and a fixed azimuthal mode $\ell=150$, for the materials under consideration. These wavefunctions correspond to the resonant wavelengths that intersect the vertical dashed line $\ell=150$ shown in Fig.~\ref{fig:lamda_res}.  We can see that the radial shape of the three wavefunctions are almost identical because the wave must reflect a similar number of times $\ell$. Thus, the radial functions for the three materials are very similar even though the resonant wavelengths are different because of the different refractive index values. 

On the other hand, if we consider a fixed wavelength, for instance $\lambda=2\,\mu$m, the radial part of the wavefunctions $F_\ell$ is expected to change considerably from one material to another, because in this case the fixed resonant wavelength corresponds to different values of $\ell$. This is shown in Fig.~\ref{fig:frb}, where the wavefunctions correspond to the wavelengths $\lambda_{TE,\ell}^{25\mu\rm{m},1}$
that intersect the horizontal dashed line plotted in Fig.~\ref{fig:lamda_res}. Observe that, in this case, both fused silica and calcium fluoride have similar wavefunctions with $\ell=104$ and $\ell=103$, respectively, while chalcogenide has a higher value of $\ell=208$. 

It is worthy to observe that the radial functions have a ``tail" that extends beyond $r=25\,\mu$m. This tail corresponds to the electromagnetic evanescent field that radiates outside the sphere, so the larger the tail the bigger the leakage, or radiative loss, of the WGM. In particular, for the case of $\lambda=2\,\mu$m, the evanescent field for chalcogenide is much smaller than the ones obtained for silica and calcium fluoride, suggesting that the radiative losses will be lower for the former material. This is discussed extensively in the next section. 

\begin{figure}[ht]
\includegraphics[width=0.47\textwidth]{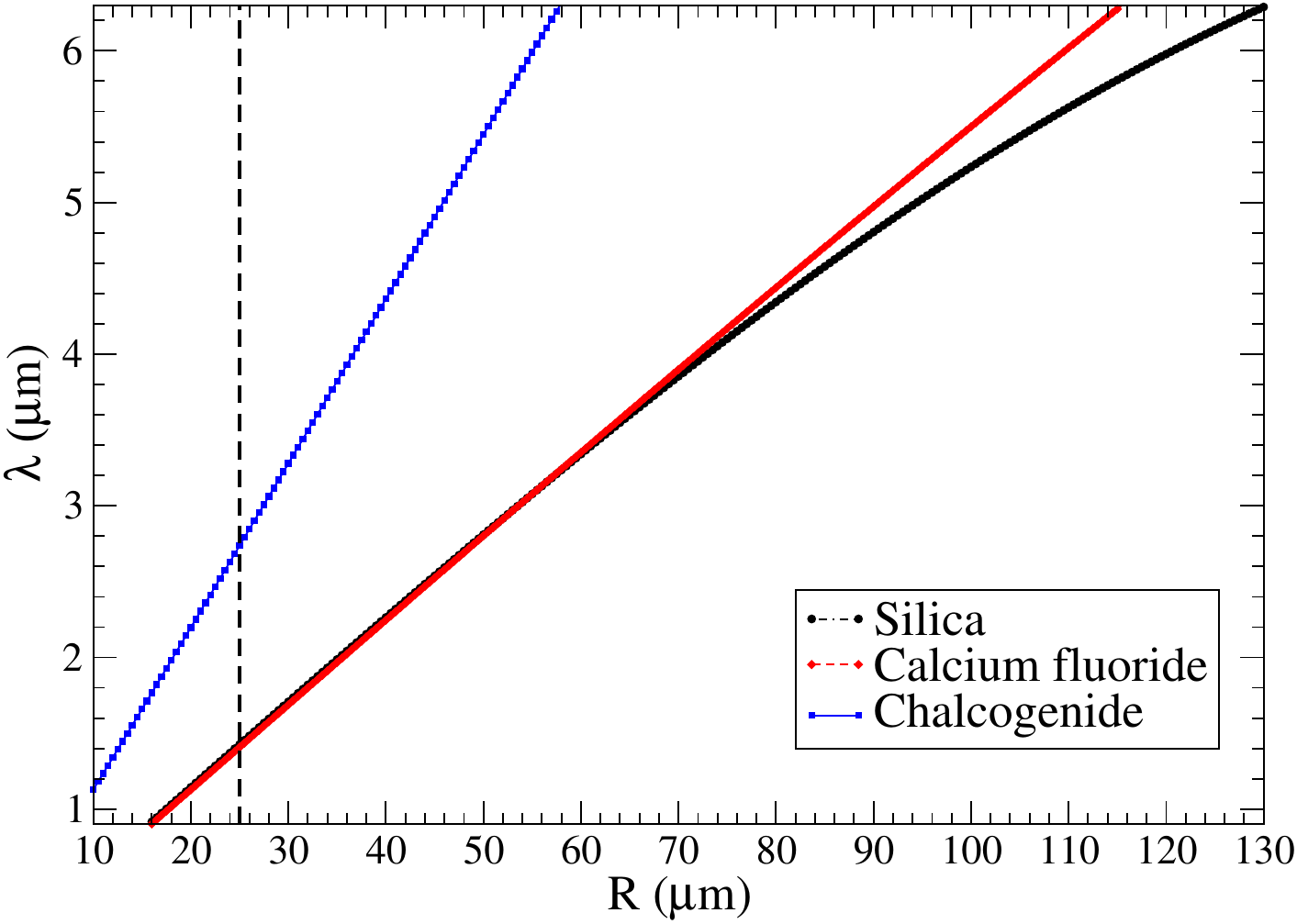}
\caption{Resonant wavelengths of the first radial mode as a function of the sphere radius, for a fixed azimuthal number $\ell=150$, for the three materials under consideration.}
\label{fig:lambda_R}
\end{figure}

We finish this section by computing the resonant WGM wavelengths for a fixed value of $\ell$, particularly $\ell=150$, for different values of the radius of the sphere. The results are shown in Fig.~\ref{fig:lambda_R}.
As discussed in~\cite{Velazquez:24}, the relevance of computing $\lambda_{TE,\nu_\ell}^{R,q=1}$ as a nonlinear equation including the dependence of the refractive index on $\lambda$, avoids the scaling of the wavelengths as a function of the radius of the sphere $R$. As it can be noticed in Fig.~\ref{fig:lambda_R}, for some materials it may well be the case that the resonant wavelengths increase linearly as a function of the radius, as it is shown for AsSe chalcogenide. But this linearity is broken for calcium fluoride and more clearly for silica for larger wavelengths.

\section{Some properties of WGM for sensing techniques}\label{sec:WGMprop}

Sensing devices can be implemented using WGM microresonators which can be designed to detect, for example: proteins, viruses, nanoparticles, temperature changes, pressure changes, or electrical and magnetic fields. Since there are different sensing mechanisms, the design of the microresonator has to be in accordance with the desired mechanism.
One of the most common sensing mechanisms is mode shifting (MS). MS sensing is usually implemented by looking for changes in the resonant wavelength of WGMs that are produced, for instance, by refractive index changes surrounding the sphere due to changes in physical parameters such as temperature, pressure, magnetic fields, among others. 
The success of spherical WGM as MS sensors can be attributed to two factors: since the acquisition time of mode shifting data is of order of tens of milliseconds, the electromagnetic mode should be confined for that amount of time. This is exactly the case for WGM modes. . 
A physical parameter that is directly related to the amount of time a WGM mode spends circulating inside the resonator is proportional to the so called {\it quality factor}, that we discuss next. 

\subsection{$Q$-factor}

\begin{figure*}[ht]
\captionsetup[subfigure]{labelformat=nocaption}
\centering
\includegraphics[width=0.5\textwidth]{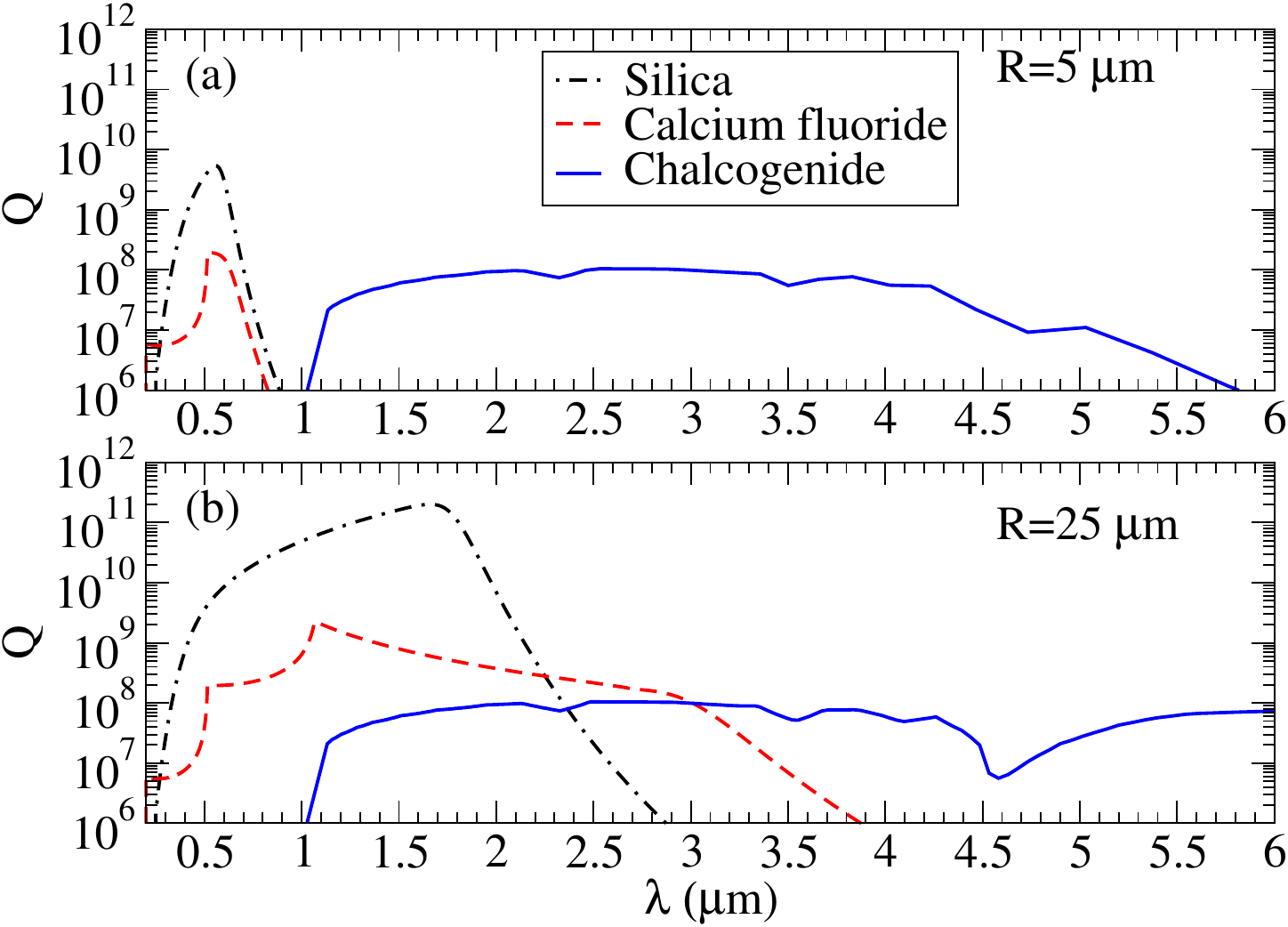}%
\includegraphics[width=0.5\textwidth]{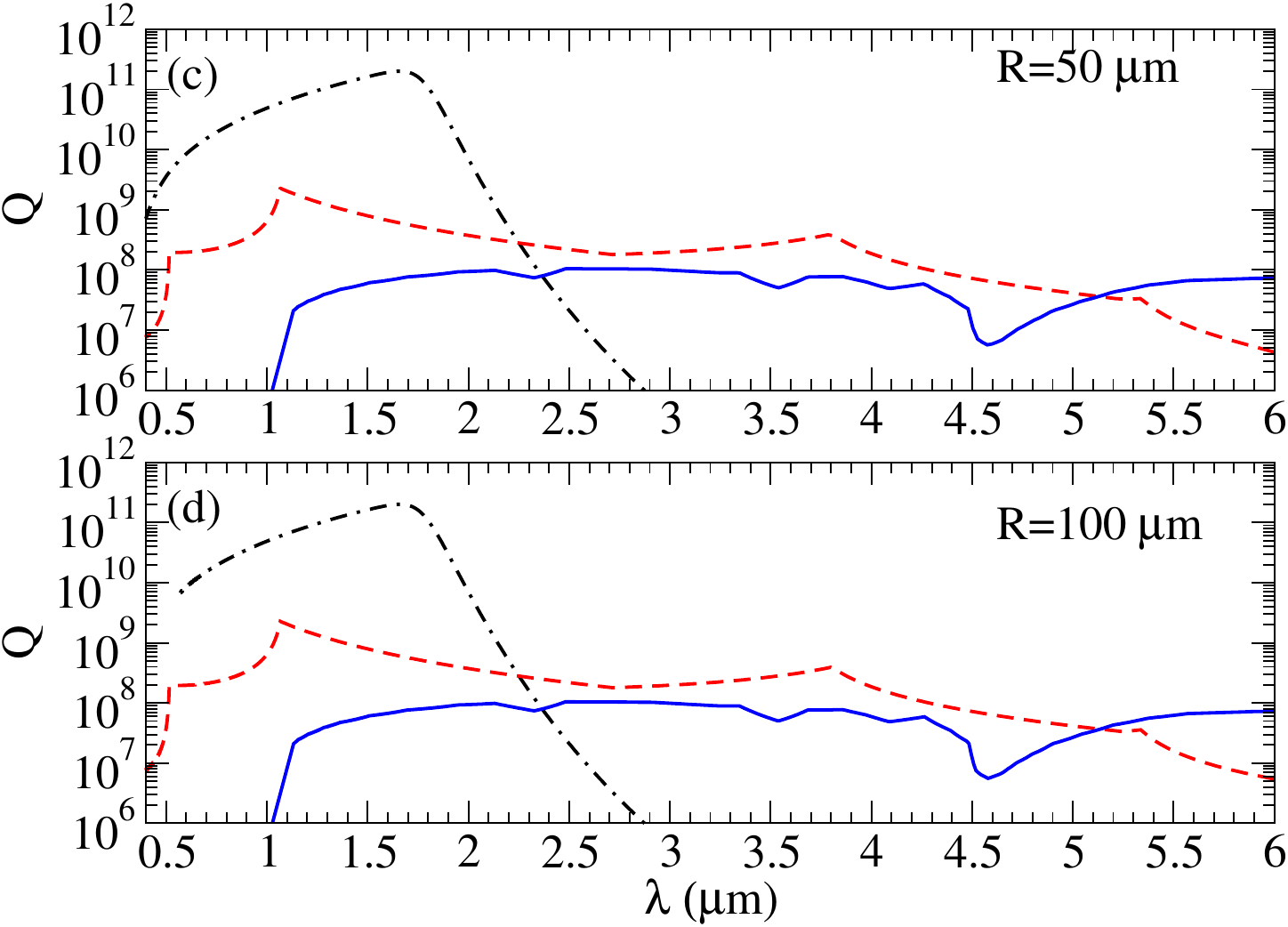}%
    \begin{subfigure}{0\linewidth}
    \caption{}\label{fig:Qa}
    \end{subfigure}
    \begin{subfigure}{0\linewidth}
    \caption{}\label{fig:Qb}
    \end{subfigure}
    \begin{subfigure}{0\linewidth}
    \caption{}\label{fig:Qc}
    \end{subfigure}
    \begin{subfigure}{0\linewidth}
    \caption{}\label{fig:Qd}
    \end{subfigure}
 \caption{$Q$-factor as a function of wavelength for fused silica (black dot-dashed line), calcium fluoride (red dashed line) and AsSe chalcogenide (blue solid line), for different sphere radii: (a)~$R=5\,\mu$m, (b)~$R=25\,\mu$m, (c)~$R=50\,\mu$m, (d)~$R=100\,\mu$m.}
    \label{fig:Q}
\end{figure*}

The intrinsic losses in a WGM resonator can be characterized in terms of the quality factor, or $Q$-factor. These losses have different contributions, such that the $Q$-factor can be expressed as

\begin{equation}
\frac{1}{Q}=\frac{1}{Q_{rad}}+\frac{1}{Q_{bulk}}+\frac{1}{Q_{ss}}+\frac{1}{Q_{W}}...\,,
\end{equation}
where we have included, for instance, the radiative loss, $Q^{-1}_{rad}$, 
which is inherent since there is an exponential attenuation of the electromagnetic field with time, so the radial contribution to the field outside the sphere is a radiation mode.
Additionally, there are other loss mechanisms that come from material non-ideal properties. 
For instance: scattering losses from residual surface inhomogeneities, $Q_{ss}$; energy losses due to absorption of the sphere material, $Q_{bulk}$; absorption due to water on the surface of the sphere, $Q_{W}$, among others.

Let us first consider theoretically the radiation leakage associated with $Q_{rad}$, which can be computed with knowledge of the resonant WGM frequencies, 
$\omega_{WGM}=\omega^{Re}+i \omega^{Im}$, as
\begin{equation}
    Q_{rad}=\frac{1}{2}\frac{\omega^{Re}}{\omega^{Im}}.\label{eq:defqrad}
\end{equation}
For TE modes, the real part $\omega^{Re}$ can be computed using Eq.~\eqref{eq:aproximated}. 
Ignoring powers of $\nu_\ell^{-2/3}$, we have
\begin{equation}
    \omega_{TE}^{Re}=\frac{c}{nR}\left(\nu_\ell-t_q^0\left(\frac{\nu_\ell}{2}\right)^{1/3}-\frac{n}{\sqrt{n^2-1}}\right)\,,\label{eq:wr}
\end{equation}
and the imaginary part $\omega^{Im}$ has been estimated in~\cite{Datsyuk:92} as 
\begin{equation}
    \omega_{TE}^{Im}=\frac{c}{R}\frac{\exp{(2T_\ell)}}{\sqrt{n^2-1}}\,,\label{eq:wi}
\end{equation}
where 
\begin{eqnarray}
 &&\exp(2T_\ell)=\exp\left[-\frac{\nu_\ell}{\beta_\ell}\sqrt{\beta_\ell^2-1}\right]\nonumber\\
 &\times&\exp\left[2\nu_\ell \ln\left(\beta_\ell+\sqrt{\beta_\ell^2-1}\right)\right].
\end{eqnarray}
Here,
\begin{equation}
\beta_\ell=\frac{n}{1-(\nu_\ell)^{-1}\left(t^0_n\left(\frac{\nu_\ell}{2}\right)^{1/3}+\frac{n}{\sqrt{n^2-1}}\right)}\,.
\end{equation}
Inserting Eqs.~\eqref{eq:wr} and~\eqref{eq:wi} into Eq.~\eqref{eq:defqrad}
we get
\begin{eqnarray}
\label{eq:qrad2}
Q_{rad}&=&\exp(2T_\ell)\times \\
&&\biggr[\left(\nu_\ell-t_q^0\left(\frac{\nu_\ell}{2}\right)^{1/3}\right)\frac{\sqrt{n^2-1}}{2n}-\frac{1}{2}\biggl]\,.\nonumber
\end{eqnarray}
$Q_{rad}$ increases very fast, as it has an exponential dependence in $\ell$, so then losses due to intrinsic radiative processes are negligible for large values of $\ell$. 

We neglect the contribution of $Q_{W}$ on our estimations of $Q$, since we consider ideal conditions of a dry environment. Similarly, we neglect the contribution of $Q_{ss}$ because it is smaller in comparison to $Q_{bulk}$ \cite{Vernooy:98}. 
The material absorption losses, $Q_{bulk}^{-1}$, can be estimated in terms of the absorption coefficient, $\alpha(\lambda)$, for each material, as follows~\cite{Gorodetsky:96}:
\begin{equation}
Q_{bulk}=\frac{2\pi n}{\alpha\lambda}\,.
\label{eq:qbulk}
\end{equation}

The absorption coefficient as a function of wavelength, $\alpha(\lambda)$, can be computed for silica trough Eq.~\eqref{eq:alpha}~\cite{Gorodetsky:00}. For calcium fluoride and AsSe chalcogenide, the absorption coefficient functions are linearly interpolated from data reported in~\cite{Li:80:2,corning} and~\cite{Ballato:17, Boudebs:04}, respectively.  

With the help of Eqs.~\eqref{eq:qrad2} and~\eqref{eq:qbulk} we can compute $Q$ for the calculated resonant wavelengths. Fig.~\ref{fig:Q} shows a plot of $Q$ as a function of wavelength for four different sphere radii: $R~=~5\,\mu$m,  ~$R=25\,\mu$m, ~$R=50\,\mu$m, and ~$R=100\,\mu$m, and for the three materials under consideration in the present work.

The fall in $Q$ values for large values of $\lambda$ observed in Fig.~\ref{fig:Q} is due to the decrease in $Q_{rad}$, while the fall for small values of $\lambda$ is due to the material absorption $Q_{bulk}$. 
We can also see the role that the radius of the sphere has on the $Q$-factor. For smaller radii, the fall of $Q_{rad}$ for large $\lambda$ suppresses $Q$ in silica and calcium fluoride spheres, while this suppression affects only longer NIR wavelengths in AsSe chalcogenide spheres (see Fig.~\ref{fig:Qa}). As the radius increases, this suppression in $Q_{rad}$ contributes less, and finally for $R>25\,\mu$m, $Q\simeq Q_{Bulk}$ (see Figs.~\ref{fig:Qb},~\ref{fig:Qc} and~\ref{fig:Qd}). Hence, all energy losses are due to material absorption. The decrease in radiation losses can be understood with the help of Fig.~\ref{fig:fr}: $\ell$ increases and the wave is confined closer to the surface and the tail of $F_\ell(R)$ for $r>R$ is smaller. Thus, higher values of $\ell$ produce an exponential increase in $Q_{rad}$ (see Eq.~\eqref{eq:qrad2}), so $1/Q_{rad} \to 0$.  
Furthermore, it is clear from Fig.~\ref{fig:Q} that for wavelengths in the $2\,\mu$m window the silica spheres present the highest Q-factor. 

\subsection{Free Spectral Range}
When considering WGM microresonators as sensors, it is desirable to have a high limit of detection (LOD). LOD is the lowest signal to be extracted that can be observed with a sufficient degree of statistical significance. 
A parameter directly related to LOD in WGM microresonators is the free spectral range (FSR). The FSR is an essential parameter involved in the design of WGM microresonators, related to the difference between two adjacent resonances. FSR is usually given in terms of the shift in frequency
spacing of two modes with the same values of $q$ but different consecutive azimuthal numbers $l$~\cite{Chiasera:10},
\begin{equation}
\Delta \omega_{FSR}=\omega_{TE,\ell+1}^{R,q}-\omega_{TE,\ell}^{R,q}.
\end{equation}
Similarly, the FSR can also be estimated as a function of wavelength.

From now on, we consider the free spectral range as computed by
\begin{equation}
\Delta \lambda_{\ell}^{FSR}=|\lambda_{TE,\ell+1}^{R,q}-\lambda_{TE,\ell}^{R,q}|.
\end{equation}
\begin{figure}[ht]
\captionsetup[subfigure]{labelformat=nocaption}
\centering
\includegraphics[width=0.48\textwidth]{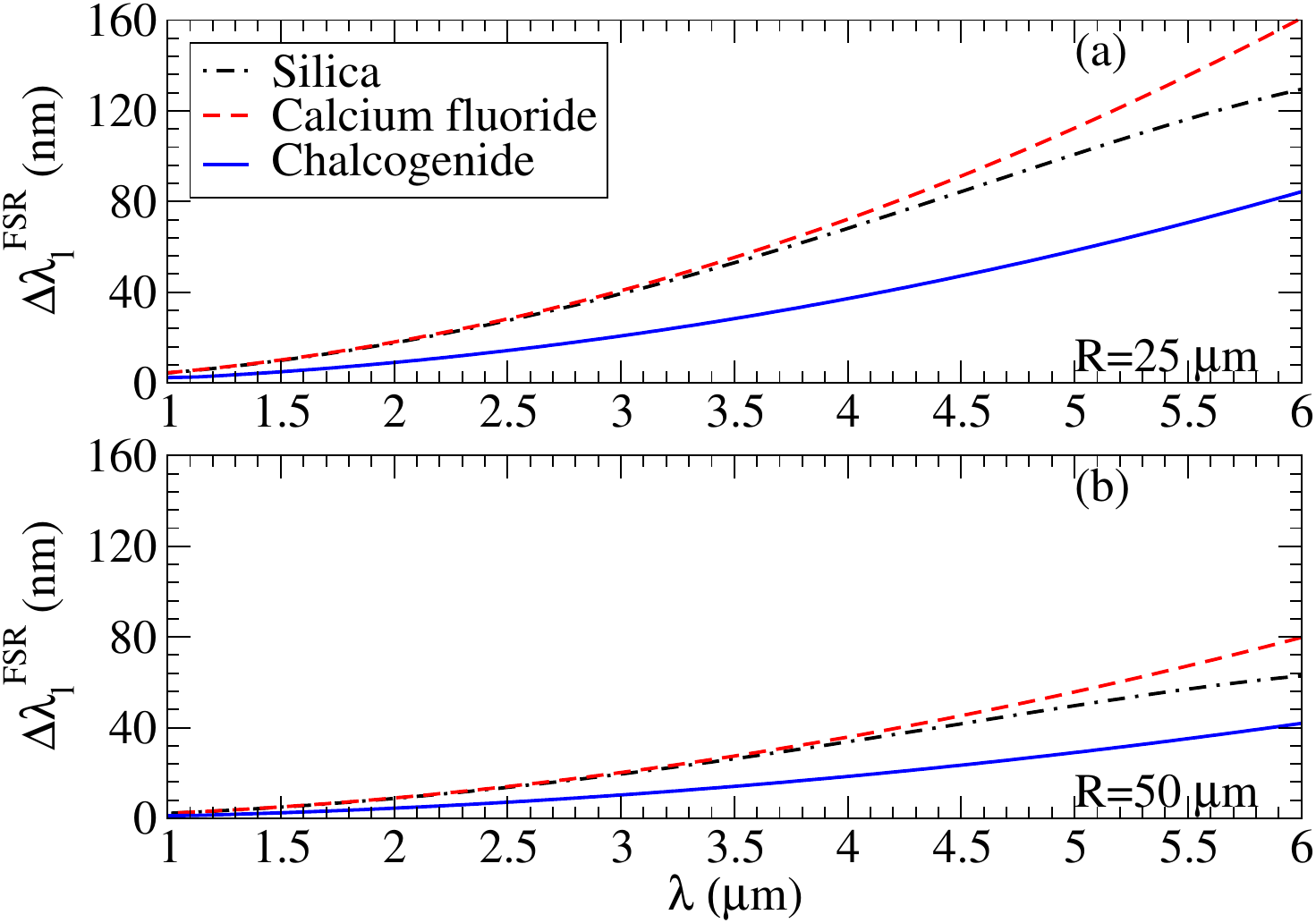}%
    \begin{subfigure}{0\linewidth}
    \caption{}\label{fig:FSRa}
    \end{subfigure}
    \begin{subfigure}{0\linewidth}
    \caption{}\label{fig:FSRb}
    \end{subfigure}
    \begin{subfigure}{0\linewidth}
    \caption{}\label{fig:FSRc}
    \end{subfigure}
    \begin{subfigure}{0\linewidth}
    \caption{}\label{fig:FSRd}
    \end{subfigure}
 \caption{$\Delta \lambda_{\ell}^{FSR}$ for the three materials under consideration, for: (a) $R=25\,\mu$m and (b) $R=50\,\mu$m. Black dot-dashed line corresponds to silica, red dashed line to calcium fluoride, and blue solid line to AsSe chalcogenide.}
    \label{fig:FSR}
\end{figure}

Fig.~\ref{fig:FSR} shows $\Delta \lambda_\ell^{FSR}$ as a function of wavelength for the three materials under consideration: Fig.~\ref{fig:FSRa} for spheres with $R=25\,\mu$m, and Fig.~\ref{fig:FSRb} for $R=50\,\mu$m.  The observed behavior can be understood if we approximate $\Delta \lambda_{\ell}^{FSR}$ using Eq.~\eqref{eq:aproximated}; at zero order we get the following expression:
\begin{equation}
    \Delta \lambda_{\ell}^{FSR} \simeq \frac{\lambda^2}{2 \pi n R}\,.\label{eq:deltal_aprox}
\end{equation}
Since the refractive index $n$ of AsSe chalcogenide is higher than that of silica and calcium fluoride, then $\Delta \lambda_{\ell}^{FSR}$ is smaller for chalcogenide than for the other two materials. 
A similar behavior results from an analogous analysis in terms of the radius of the sphere $R$. 

Other types of FSR can be defined; for instance, as the difference between TE and TM modes with the same $\ell$ and $q$; or the difference between the resonances of two adjacent radial modes, i.e. for two consecutive values of $q$, 
\begin{equation}
\Delta \lambda_{q}^{FSR}=\lambda_{TE,\ell}^{R,q+1}-\lambda_{TE,\ell}^{R,q}\,.
\end{equation}

As before, we estimate $\Delta \lambda_{q}^{FSR}$ to first order with the use of Eq.~\eqref{eq:aproximated},
\begin{equation}
    \Delta \lambda_{q}^{FSR} \simeq \left(\frac{2^2 \lambda^5}{\pi n^2R^2}\right)^{1/3}\left(t_q^0-t_{q+1}^0\right)\,.\label{eq:deltaq_aprox}
\end{equation}
A further approximation valid for $\ell \gg 1$ in Eq.~\eqref{eq:deltal_aprox} reveals that 
\begin{equation}
 \Delta \lambda_{\ell}^{FSR} \sim \frac{nR}{\ell^2},
\end{equation}
and similarly for Eq.~\eqref{eq:deltaq_aprox} 
\begin{equation}
\Delta \lambda_{q}^{FSR}\sim \frac{nR}{\ell^{5/3}}.
\end{equation}
Thus,  from this analysis it is evident that $\Delta \lambda_{\ell}^{FSR} \ll \Delta \lambda_{q}^{FSR}$ for $\ell\gg 1$. 

The sensitivity of an optical sensor is usually defined as the ratio of signal changes from the sensor to the variation in the measured parameter.
A good sensor based on WGM resonators is linked with the avoidance of a true mode shifting due to the measured quantity of interest with the presence of another WGM resonance. Thus, a large FSR and a small true mode shift are desirable. 
Since $\Delta \lambda_{\ell}^{FSR}$ is the smallest of the material proper resonances, we only have to worry to be below this FSR when comparing the mode shifting induced by the quantity under measurement.  
Based on the previous discussion, Fig.~\ref{fig:FSR} gives some insight about which material under the present study could potentially be better for sensing applications.
For NIR wavelengths, calcium fluoride has a slightly higher value of $\Delta \lambda_{\ell}^{FSR}$ than fused silica, independently of the radius of the sphere. 
On the other hand, AsSe chalcogenide is the material with lowest $\Delta \lambda_{\ell}^{FSR}$ for all radii.

In the design of sensors based on WGM resonators, in addition to the absolute magnitude of $\Delta\lambda_\ell^{FSR}$, it could be useful to known the FSR response to any environmental changes
that influence the refractive index of the sphere. In order to see how FSR is affected by changes in $n$, we define 
a ``derivative" of $\Delta \lambda_{\ell}^{FSR}$ as follows: 
\begin{equation}
\frac{\delta \Delta \lambda_\ell^{FSR}}{\delta n}=\frac{\Delta\lambda_{\ell+1}^{FSR}-\Delta\lambda_\ell^{FSR}}{n(\lambda_{\ell+1})-n(\lambda_\ell)}\,.\label{eq:derivada}
\end{equation}
which can be interpreted as the susceptibility of FSR to changes in $n$. 

\begin{figure}[ht]
 \includegraphics[width=0.5\textwidth]{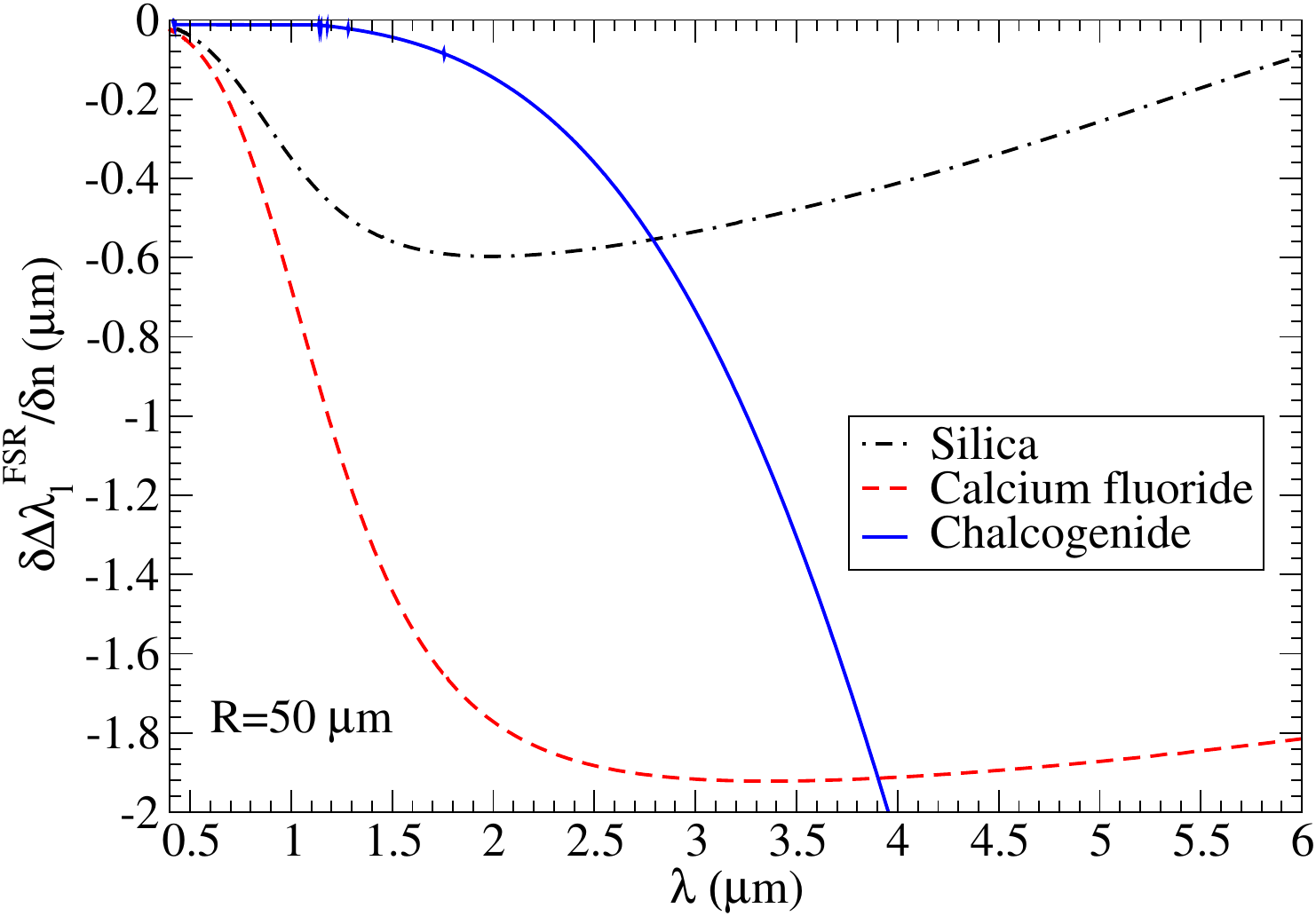}
			\caption{Changes in the FSR due to changes in the refractive index as a function of wavelength for $R=50\,\mu$m, estimated by Eq.~\eqref{eq:derivada}. }
			\label{fig:dFSR}
\end{figure}

We have plotted $\frac{\delta \Delta \lambda_\ell^{FSR}}{\delta n}$ for a sphere with radius $R=50\,\mu$m in  Fig.~\ref{fig:dFSR}, for the three materials under consideration.  

Observe that CaF$_2$ has the largest sensitivity (in magnitude) to changes in the refractive index, while AsSe chalcogenide has the lowest at the $2\,\mu$m window. If the change in $n$ is produced by effect of the environment that can be understood as a source of noise in sensing experiments, thus we can conclude from this analysis that AsSe chalcogenide is the material with less source of error.

\section{Discussion and conclusions}\label{sec:disc}

\begin{figure}[ht]
 \includegraphics[width=0.45\textwidth]{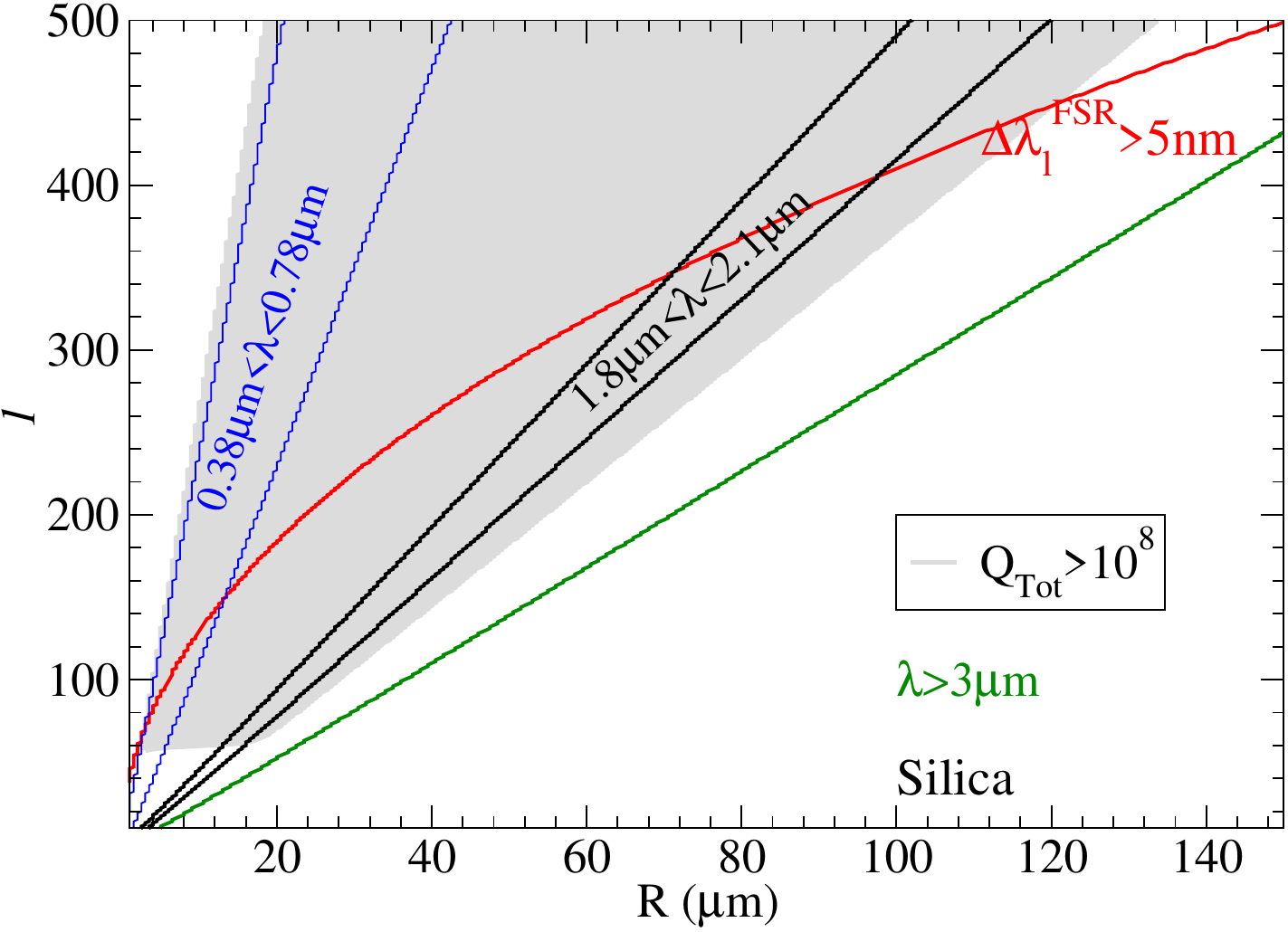}
 \includegraphics[width=0.45\textwidth]{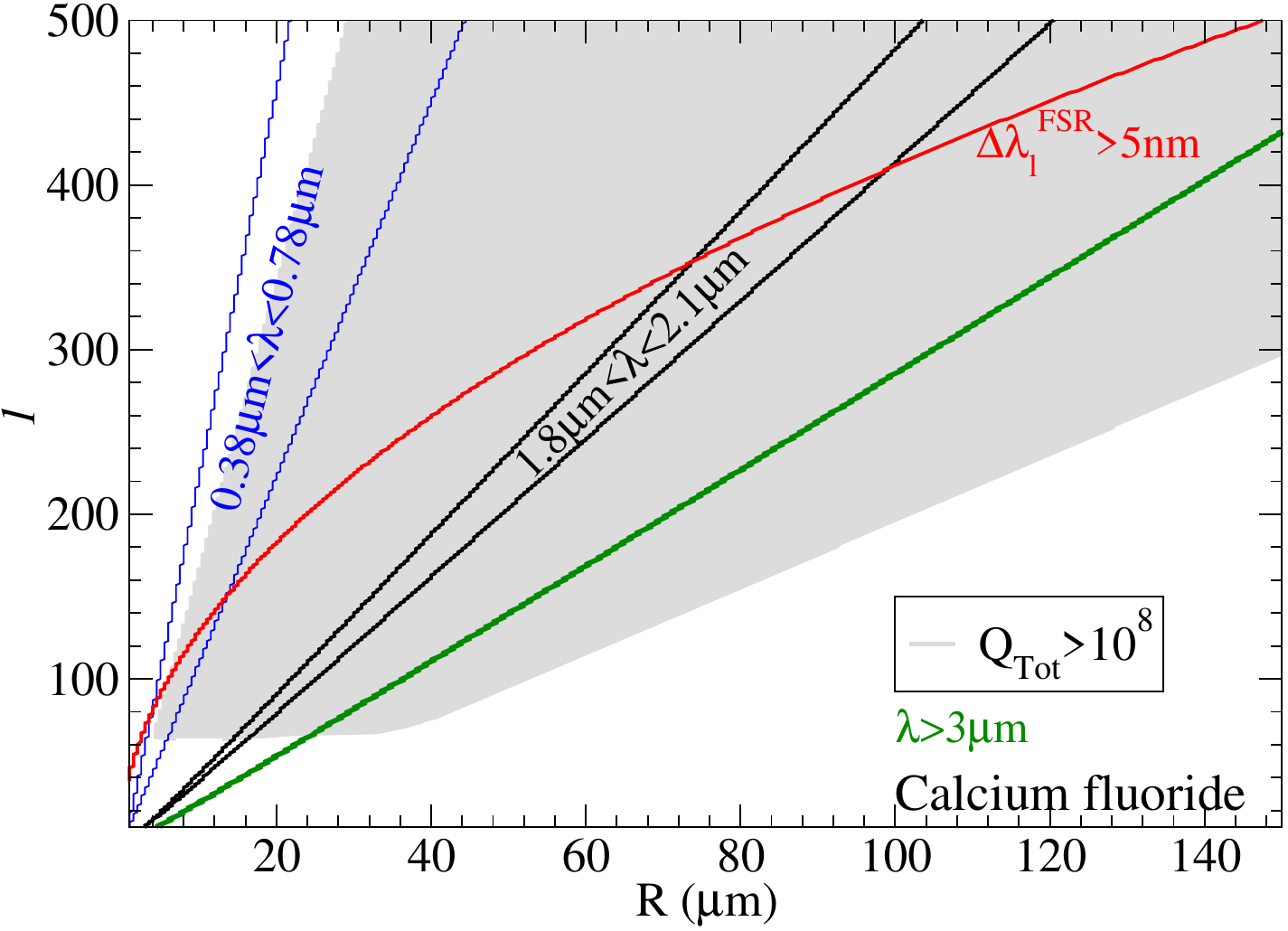}
 \includegraphics[width=0.45\textwidth]{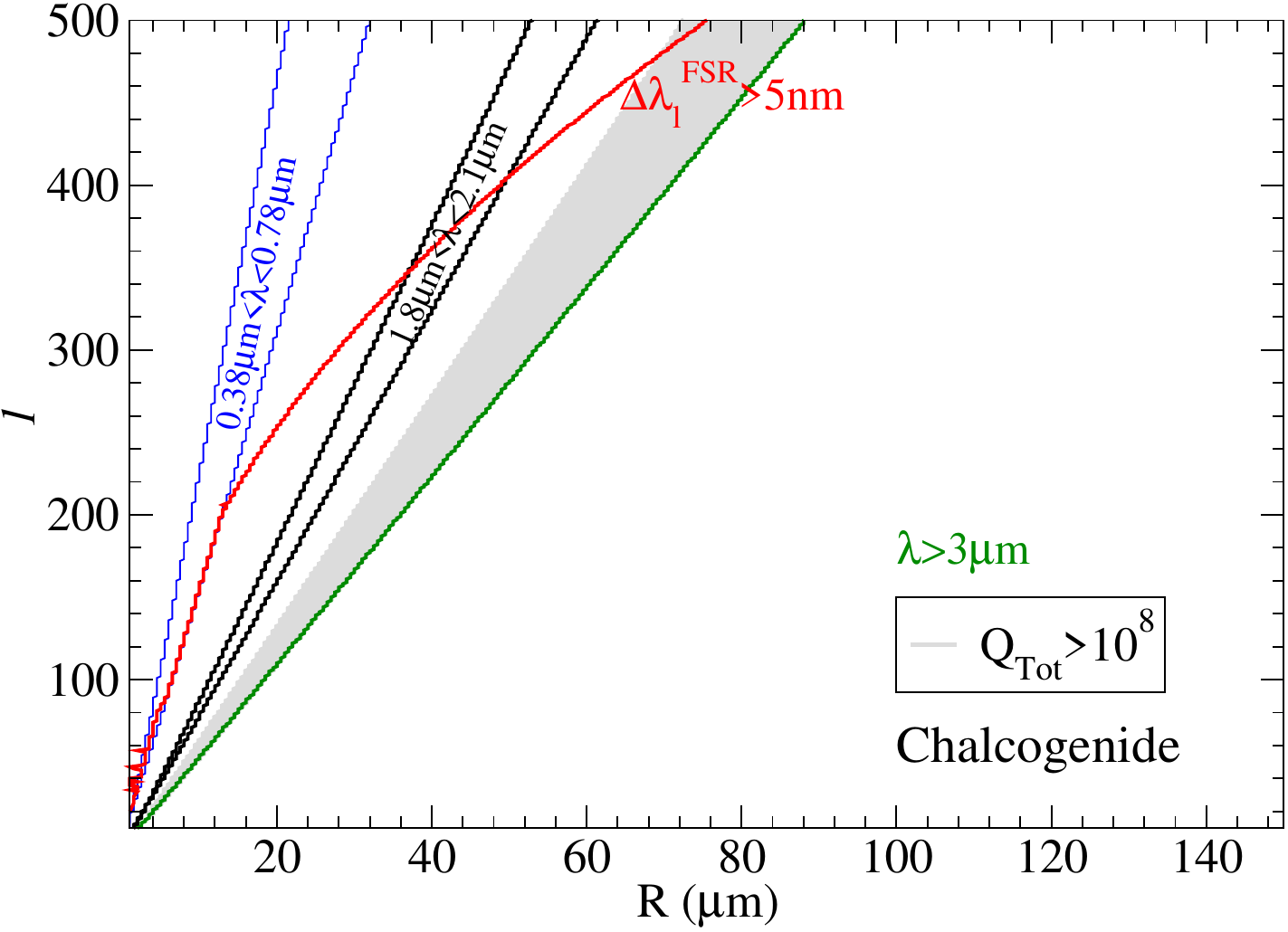}
			\caption{Summary of the main properties that optimize microspheres as WGM mode shifting sensors. See text for explanation.}
			\label{fig:all}
\end{figure}    
Optical WGM spherical microresonators have been successfully used as sensors mainly because of their high $Q$-factor, i.e. a large time resolution. Furthermore, their high $\Delta \lambda_\ell^{FSR}$ implies a high detection limit. These two properties are intrinsically related with the radius of the resonator and the physical properties of the material used for their fabrication.
Because of the imminent saturation in the optical and telecommunication wavelengths bands, the wavelength region around $2\,\mu$m has attracted significant interest. 
In the present work, we have computed the $Q$-factor and the free spectral range of microspheres of different materials and radii in order to seek for the optimal radius and material that maximize both $Q$-factor and $\Delta \lambda_{\ell}^{FSR}$.

Our findings are summarized in Fig.~\ref{fig:all}, where we have plotted in the space parameter $(R,\ell)$ isocurves of the resonant wavelengths 
$\lambda_{TE,\ell}^{R,1}$ that correspond to the wavelength ranges: $0.3\,\mu\mbox{m} <\lambda_{TE,\ell}^{R,1} <0.75\,\mu\mbox{m}$ (blue lines), i.e. the optical band,  and $1.8\,\mu\mbox{m} <\lambda_{TE,\ell}^{R,1} <2.1 \,\mu\mbox{m}$ (black lines), which correspond to what is now called the $2\,\mu$m window. Below the green line are the resonances that correspond to $\lambda_{TE,\ell}^{R,1} \ge 3.0\,\mu\mbox{m}$. 
Another isocurve is shown in a red line, that corresponds to $\Delta\lambda_\ell^{FSR}=5$\,nm.  Above this isocurve, spheres with a radius as given by the horizontal axis have WGM resonances with FSR values lower than 5\,nm, while below this line the WGM resonances have FSR values greater than 5\,nm. Finally, the grey region corresponds to the WGM resonances for the first radial mode that have $Q$-factor values $Q>10^{8}$. 

In summary, we have found that WGM sensors can be optimized for microspheres with radius in the range $20\,\mu\mbox{m}<R<100\,\mu\mbox{m}$ for silica and calcium fluoride spheres, while in the case of AsSe chalcogenide, spheres with $R<40\,\mu\mbox{m}$ are desirable for the $2\,\mu$m window.
The lower limit in $R$ comes from the requirement that $Q \gg 1$, while the upper bound on $R$ comes from the requirement of a large FSR. To understand this upper bound, we refer to Eq.~\eqref{eq:deltal_aprox} where we can see a competition between a quadratic dependence of $\Delta \lambda_\ell^{FSR}$ with wavelength $\lambda$ and an inverse dependence in $R$ that limits the size of the spheres in the optimization of $\Delta \lambda_\ell^{FSR}$.

The presented methodology can be extended to other geometries and materials, making it suitable for studying the optimal design of WGM resonators across different applications.

\bmhead{Acknowledgements}
This work was partially supported by CONAHCYT-SNII.

\bibliographystyle{unsrt}
\bibliography{biblio}


\end{document}